 \documentclass{myaa}    
\usepackage{graphicx}  
\usepackage{rotating}  

\def\beq{\begin{equation}}  
\def\eeq{\end{equation}}

\begin{document}  
    \title{The Nuclear Region of \\   
     Low Luminosity Flat Radio Spectrum Sources\thanks{Based on   
     observations collected at the Multiple Mirror Telescope  
     on Mt. Hopkins.}}  
      
\subtitle{I. Stellar Content}  
  
    \author{M. Serote Roos\inst{1} \and A.\,C. Gon\c{c}alves\inst{1,2}}  
  
    \offprints{M. Serote Roos}  
  
\institute{Centro de Astronomia e Astrof\'{\i}sica da Universidade de Lisboa,  
Observat\'orio Astron\'omico de Lisboa,  Tapada da Ajuda, P-1349-018 Lisboa,  
Portugal \\  
\email{serote@oal.ul.pt, adarbon@oal.ul.pt}  
\and  
European Southern Observatory, Karl-Schwarzschild-Strasse 2,  
D-85748 Garching bei M\"unchen, Germany \\  
\email{adarbon@eso.org}}  
  
  \date{Received 11 December 2001 / Accepted 4 September 2003}  
  
\abstract{  
In this work we have examined the  spectroscopic properties of  a   
sample of 19 optically bright, low luminosity   
Flat Radio Spectrum (LL FRS) sources. Our study focuses on the properties   
of their host galaxies, namely the nuclear stellar populations   
and dust content.   
In the optical --- spectral region covered by our data --- the objects   
in the sample are mainly dominated by the host galaxy starlight,   
which strongly dilutes the non-thermal continuum as well as possible   
emission-line features related to the active nucleus.  
We have computed the nuclear stellar populations contributing   
to the spectra of the objects in our sample. The stellar population     
synthesis has been performed by using a very reliable mathematical   
method, which yields a Global Principal Geometrical solution.   
Our results show that, for most of the objects in the sample, the   
populations are composed of old stars of solar metallicity, or   
lower; the populations are mainly composed of late-type stars,   
i.e. G, K and M spectral types, the young component coming thus   
from supergiant stars; the dust content is weak. Both the stellar   
populations and the dust content are in agreement with what is   
usually observed in ``normal''  elliptical galaxies. Similar stellar   
content  has equally been found in the nuclear regions of galaxies   
hosting a Low Ionization Nuclear Emission Line Region, or LINER.   
  
The present work is important in illustrating the different applications   
of stellar population synthesis in the study of low luminosity   
radio sources. In fact, the synthesis allows us not only to   
obtain valuable information about the stellar populations and dust   
content of the host galaxies, therefore providing material for   
further studies on the connection between host galaxy and active   
nucleus, but also to reveal the so-far unstudied optical emission-line   
features present in the spectrum of our objects.   

}  
  \titlerunning{Nuclear Region of LL FRS Sources: Stellar Content}  
  \authorrunning{Serote Roos \& Gon\c{c}alves}  
  \maketitle  
  
\section{Introduction}  
BL Lacertae objects (BL~Lacs), together with Flat Spectrum Radio   
Quasars (FSRQs) are called blazars. Blazars display some of the   
most extreme behaviour observed in Active Galactic Nuclei (AGN),   
such as high and variable optical polarization; they also show a   
strong and rapid variability at radio, optical and X-ray wavelengths.   
BL Lacs  usually have weak or non-observed optical emission lines.    
  
Blazars  exhibit characteristics indicative of  
relativistic beaming (Padovani \& Urry 1990), which occurs when the   
relativistic emission (a jet) is oriented close to the   
observer's line of sight (Blandford \& Rees 1978). The extreme   
properties of blazars are generally interpreted as a consequence   
of non-thermal emission from the relativistic jet.    
A transition population between beamed BL Lacs and unbeamed radio-galaxies   
has not been detected (Rector et al. 1999); yet, a transition population of   
``low luminosity BL Lacs'' (LL BLLs) was predicted to exist in abundance   
in samples like the one studied by Browne \& March\~a (1993).  
  
In the past few years there has been a growing interest in   
LL BLLs and other flat radio spectrum sources.   
Because they are relatively close to us, their host galaxies can be studied   
more easily. The knowledge of their host galaxies is of the utmost   
importance, not only as a test of the unification schemes, but   
also as a way to better understand the relationship between nuclear   
activity and host galaxy. BL~Lac host galaxies seem to be luminous   
ellipticals, often in loose groups; in fact, from a morphological   
point of view, BL~Lac host galaxies seem to be absolutely normal   
ellipticals (Urry et al. 2000; Falomo \& Ulrich 2000). {\it HST}   
observations by Verdoes Kleijn et al. (1999) suggest that, apart   
from the presence of dust and gas distributions, the nuclei of   
FR~I radio-galaxies also resemble those of normal bright ellipticals.   
While there are some studies concerning the host galaxy morphology, little   
is, however, known about the nuclear stellar populations of Low   
Luminosity Flat Radio Spectrum (LL FRS) sources.   
  
In an effort to better understand such objects and   
their host galaxies, we undertook the study of the optical   
properties of a sample of low luminosity, core-dominated radio   
sources expected to contain a high fraction of objects observed  
at small angles to the line-of-sight. This sample is known to   
contain several weak emission line galaxies, which could be objects  
related to BL~Lac phenomena, but observed at larger angles from the   
line-of-sight; it is also possible that some of the galaxies  
are ``hidden'' BL Lacs, whose nuclear emission is completely swamped   
by the host galaxy starlight (Dennett-Thorpe \& March\~a 2000).   
  
\vspace{0.5mm}  
With this study, we had two main purposes:    
  
{\it (i)} studying the nuclear stellar populations   
of low luminosity, flat radio spectrum nuclei such as LL BLLs;    
  
{\it (ii)} to search for the clear optical signature of the   
underlying activity.   
  
\vspace{0.5mm}  
To do this, we must rely on the computation of the nuclear stellar   
populations with subsequent subtraction of their contribution to the   
total nuclear spectra.  Our stellar synthesis is not done   
by using templates of normal nearby galaxies to estimate the stellar   
populations. In fact, we take the spectra of our objects to compute   
the stellar populations; in   
this way, the information on the stars that compose our galaxies   
is taken directly from the data, without any {\it a priori}   
assumption on the nature of the stellar populations.  

\vspace{0.5mm}  
In Sect.~2 we describe the sample from which   
we have selected the objects discussed in this paper; we present   
our own sub-sample, as well as the criteria behind its selection.   
Section~3 briefly describes the observational data analyzed in   
the paper. In Sect.~4 we introduce the stellar population synthesis   
method; we also describe how it has been applied to the data,   
presenting the results obtained for each galaxy.   
In Sect.~5 we summarize our results and conclusions.
  
\section{The sample}  
Being interested in studying the optical properties  of LL FRS   
sources, we undertook the analysis of a set of such objects taken   
from the sample discussed in March\~a et al. (1996).  This sample   
presents objects with a redshift up to  0.1. Since it only contains     
nearby objects, the host galaxies are therefore easy to study.   
  
\begin{table*}[t]  
\caption{\label{sample} List of the 19 objects selected from  
the 200~mJy sample for this study.  
In Col.~1 we give the identification of the objects based on their  
B1950 radio positions and in Col.~2 some of the names under which  
their counterparts are known; in Cols.~3 and 4 we give the objects'  
coordinates and in Col.~5 the redshift; Cols. 6, 7, 8  and 9  
correspond to the contrast values$^{2}$ and to the radio (8.4 GHz)  
and optical polarization, respectively.  
Values in Cols. 6 and 9 were taken from March\~a et al. (1996);  
the data given in Cols. 7 and 8 are from Dennett-Thorpe \&  
March\~a (2000).}  
\begin{center}  
\begin{tabular}{clcccccrr}  
\hline  
\hline  
ID & Name& $\alpha$ (J2000) & $\delta$ (J2000) & $z$ & $C_{\lambda}$  
&  $C_{\nu}$ & $P_{\rm r}$ (\%) &  $P_{\rm o}$ (\%) \\  
\hline  
0035+227 & MG3 J003808$+$2303  
& 00 38 08.1 & 23 03 28.4 & 0.097 & 0.44 & 0.57 & $<$0.81 & 0.84 \\  
  
0055+300 & NGC 315  
& 00 57 48.9 & 30 21 08.8 & 0.017 & 0.50 & 0.56 & $<$0.18 & --~~ \\  
  
0116+319 & 4C 31.04  
& 01 19 35.0 & 32 10 50.0 & 0.060 & 0.46 & 0.56 & 0.22 & $<$1.15 \\  
  
0149+710 & RX J0153.3$+$7115  
& 01 53 25.8 & 71 15 06.5 & 0.023 & 0.32 & 0.46 & 2.67 &  3.31 \\  
  
0651+410 & Zw 204$-$27  
& 06 55 10.0 & 41 00 10.1 & 0.022 & 0.47 & 0.55 & $<$0.43 &  0.47 \\  
  
0729+562 & TXS 0729$+$562  
& 07 33 28.6 & 56 05 41.7 & 0.107 & 0.43 & 0.61 & $<$1.02 & $<$1.11 \\  
  
0733+597 & RX J0737.4$+$5941  
& 07 37 30.1 & 59 41 03.2 & 0.041 & 0.43 & 0.58 & $<$0.62 & $<$0.70 \\  
  
0848+686 & CGCG 332$-$026  
& 08 53 18.9 & 68 28 19.0 & 0.041 & 0.46 & 0.54 & 2.54 & 0.74 \\  
  
0902+468 & TXS 0902+468  
& 09 06 15.5 & 46 36 19.0 & 0.085 & 0.48 & 0.54 & $<$0.99 & $<$1.20 \\  
  
1144+352 & RX J1147.3$+$3500  
& 11 47 22.1 & 35 01 07.5 & 0.064 & 0.43 & 0.51 & 0.44 & $<$0.63 \\  
  
1146+596 & NGC 3894  
& 11 48 50.4 & 59 24 56.4 & 0.011 & 0.50 & 0.56  & $<$0.31 & 0.36 \\  
  
1241+735 & 8C 1241$+$735  
& 12 43 11.2 & 73 15 59.3 & 0.075 & 0.43 & 0.49 & 8.02 & $<$0.75 \\  
  
1245+676 & PGC 043179  
& 12 47 33.3 & 67 23 16.5 & 0.107 & 0.50 & 0.56 & $<$1.02 & $<$1.31 \\  
  
1558+595 & PGC 056566  
& 15 59 01.7 & 59 24 21.8 & 0.060 & 0.56 & 0.57 & $<$1.06 & $<$0.78 \\  
  
1658+302 & PGC 059420  
& 17 00 45.2 & 30 08 12.9 & 0.036 & 0.45 & 0.54 & $<$1.97 & $<$0.80 \\  
  
1703+223 & 87GB 170321.7$+$221932  
& 17 05 29.3 & 22 16 07.6 & 0.050 & 0.45 & 0.53 & --~~ & 0.43 \\  
  
1755+626 & NGC 6521  
& 17 55 48.4 & 62 36 44.1 & 0.028 & 0.53 & 0.59 & 1.62 & 0.41 \\  
  
2202+363 & 87GB 220211.3$+$361757  
& 22 04 21.1 & 36 32 37.1 & 0.075 & 0.48 & 0.55 & $<$1.26 & 0.52 \\  
  
2320+203 & RX J2323.3$+$2035  
& 23 23 20.3 & 20 35 23.5 & 0.039 & 0.41 & 0.53 & 1.41 &  --~~ \\  
  
\hline  
\end{tabular}  
\end{center}  
\end{table*}  
  
March\~{a}'s et al. original sample contains 57 optically bright   
($V \leq$ 17), nearby ($z \leq$ 0.2), radio sources; these   
were selected on the basis of the spectral slope   
($\alpha _{\rm r} \leq$ 0.5 between 1.4 and 5 GHz) and are   
all located north of 20\degr, being off the galactic plane   
by at least 12\degr. The sample has a flux density limit of   
200 mJy at 5 GHz.  
  
\vspace{0.5mm}  
The main purposes of such a sample were:    
  
{\it (i)} to find LL BLLs with radio luminosities comparable to  
 X-ray selected BL~Lacs (most of which we now refer to as   
``high frequency peak'' BL Lacs);   
  
{\it (ii)} to investigate the differences between BL~Lacs and   
other flat spectrum radio sources.   
  
\vspace{0.5mm}  
Spectroscopy of these objects yielded a high fraction of BL~Lacs   
($\sim$\,35\%) and a relatively high number ($\sim$\,20\%) of   
sources with strong emission lines.  The remainder of the sample   
is composed of objects with hybrid properties (i.e.   
objects showing simultaneously BL~Lac properties, such as high   
and variable polarization, and broad emission lines, usually   
associated with the spectra of quasars) and by galaxies which   
do not show any conspicuous signs of activity, the main   
contribution to the their optical spectra being of stellar origin.    
  
In this paper we focus our attention on the latter. These galaxies, 
although radio selected  together with confirmed BL~Lacs and objects 
displaying a ``Seyfert-type'' emission spectrum, following exactly 
the same  criteria, are optically very different from them, showing   
no strong emission lines (as Seyferts do), nor a flat continuum   
and no emission or absorption features (as usually observed in BL~Lacs).   
This persuaded us that  there should be more to these sources   
than meets the eye.   
  
Among the galaxies in March\~a's sample which displayed such   
characteristics, we have selected 19 objects according to the   
following criteria:   
  
{\it (i)} the dilution of the absorption features due to the power-law  
continuum must be small enough so that information can be   
inferred from the lines. We took the 4000~\AA\ break contrast  
$C$ = 0.3\footnote{The 4000 \AA\ break contrast $C$  is defined as  
$(S^{+}-S^{-})/S^{+}$, where $S^+$ and $S^-$ are the fluxes  
measured redward and blueward of the break, respectively.   
Several tests with our population synthesis  
method have shown that for lower values of $C$ the stellar   
populations cannot be determined.} as our lowest possible limit.   
  
{\it (ii)} the signal to noise ratio must be high enough so that the   
continuum can be traced and the absorption features used.   
  
\vspace{0.5mm}  
Table~\ref{sample} gives the list of the 19 galaxies selected for  
this study; for the sake of consistency, we have kept the same  
nomenclature as in March\~a et al. (1996).  
  
\section{Observational data}  
The data consist of optical spectra ($\sim$\,3500--8700~\AA\ and  
$\sim$\,3500--7400~\AA) obtained at the Multiple Mirror Telescope   
on Mt. Hopkins.  
The spectra were acquired through a 1\farcs5 slit, oriented along  
the parallactic angle, and have a spectral resolution of 20~\AA\ (FWHM);   
they have been flux-callibrated using the spectrophotometric standards   
observed on each night. All spectra were reduced according to the standard   
procedure using IRAF and ESO/MIDAS packages. Further observational and   
reduction aspects are discussed in detail in March\~a et al. (1996).  
  
The spectra have been corrected for the interstellar reddening   
using the Howarth (1983) Galactic reddening law; the values of   
$E(B-V)$ for each galaxy were calculated using the maps of dust   
IR emission from Schlegel et al. (1998).   
A correction to the broadening of the lines due to velocity dispersion  
is not necessary, since our spectral resolution is low.  
  
\footnotetext[2]{The break contrast values given by Dennett-Thorpe   
\& March\~a (2000) are slightly different than those calculated   
by March\~a et al. (1996); this is because they use the   
$S_{\nu}$, rather than the $S_{\lambda}$ flux. The selection   
of our targets is not affected by this difference. }  
\begin{table*}[t]  
\caption{\label{EW}  
Wavelength intervals defined for the  equivalent   
widths measurement (all values in units of~\AA).}  
\begin{center}  
\begin{tabular}{lcc|lcc}  
\hline  
\hline  
Line Identification & $\lambda_{\rm central}$ & Wavelength Interval &   
Line Identification & $\lambda_{\rm central}$ & Wavelength Interval \\  
\hline  
FeI,NiI,CrI&3570&3543-3598 & FeI,TiO&5653&5630-5676\\  
  
FeI,NiI,CaI,TiI&3630&3598-3661 & FeI,NaI&5700&5676-5725\\  
  
FeI,CrI,TiI,NiI&3677&3661-3693 & FeI,TiO&5775&5725-5825\\  
  
FeI,CaII,TiI,TiII,NiI,CN&3736&3693-3780 & FeI,TiO,CaI&5850&5825-5874\\  
  
H10,CN L band&3794&3780-3808 & NaI&5894&5874-5914\\  
  
H9,CN L band,FeI,MgI,HeI&3835&3808-3862 & FeI,Ti,MnI&5972&5914-6029\\  
  
H8,CN L band,FeI,SiI,HeI&3885&3862-3908 & FeI,CaI&6070&6029-6110\\  
  
CaIIK&3930&3908-3952 & FeI&6129&6110-6148\\  
  
CaIIH,H$\epsilon$&3970&3952-3988 & TiO,CaI&6178&6148-6208\\  
  
FeI,HeI&4004&3988-4020 & TiO,FeI&6266&6208-6325\\  
  
FeI,HeI&4037&4020-4054 & FeI,CaH,TiO&6350&6325-6374\\  
  
FeI,SrII&4068&4054-4082 & FeI,CaI&6428&6374-6481\\  
  
H$\delta$&4100&4082-4118 & CaI,TiO&6508&6481-6535\\  
  
FeI&4138&4118-4159 & H$\alpha$,TiO&6558&6535-6582\\  
  
CN&4186&4159-4214 & FeI,TiO&6602&6582-6622\\  
  
CaI&4229&4214-4244 & TiO&6636&6622-6651\\  
  
FeI,CrI&4260&4244-4277 & TiO,FeI&6670&6651-6690\\  
  
CH G band,FeI,CrI&4298&4277-4318 & TiO,FeI,CaI&6726&6690-6761\\  
  
H$\gamma$,FeI,FeII&4341&4318-4364 & FeI,MgII&6778&6761-6795\\  
  
FeI,C$_2$,FeII,TiII&4392&4364-4420 & TiO,CaH&6811&6795-6827\\  
  
FeI,CaI,TiO,MgII,HeI&4446&4420-4472 & FeI,SiI&7008&6969-7048\\  
  
CH,CN&4489&4472-4506 & TiO&7063&7048-7078\\  
  
FeI,FeII,TiII&4537&4506-4568 & TiO,NiI,FeI&7109&7078-7140\\  
  
FeI,FeII,TiO,TiII,CN,CaI&4595&4568-4622 & TiO,VO&7359&7341-7377\\  
  
FeI,TiO,C$_2$&4655&4622-4688 & FeI,VO&7406&7377-7434\\  
  
FeI,MgI,TiI,HeI,NiI,C$_2$&4714&4688-4741 & FeI,VO&7458&7434-7482\\  
  
FeI,MgH,NiI,MnI,TiO&4772&4741-4802 & FeI,VO&7511&7482-7540\\  
  
TiO,MgH,CN,MnI&4818&4802-4835 & TiO,OI&7780&7737-7823\\  
  
H$\beta$,TiO,FeI&4856&4835-4876 & TiO,CN,VO&7855&7823-7888\\  
  
FeI&4886&4876-4897 & VO,CN,FeI,SiI&7929&7888-7970\\  
  
FeI,FeII,CN,HeI&4922&4897-4946 & FeI,CN&8015&7970-8060\\  
  
FeI,TiO,TiI&4972&4946-4998 & TiO,FeI&8432&8411-8453\\  
  
FeI,FeII,TiO,CN,HeI,TiI&5028&4998-5058 & TiO,FeI&8472&8453-8490\\  
  
FeI&5107&5058-5156 & CaII&8498&8490-8508\\  
  
FeI,MgI+MgH&5198&5156-5240 & FeI,VO,TiI&8518&8508-8527\\  
  
FeI&5274&5240-5308 & CaII&8542&8527-8557\\  
  
FeI&5336&5308-5356 & FeI,VO,TiO&8601&8557-8645\\  
  
FeI&5388&5356-5421 & CaII&8662&8645-8677\\  
  
FeI,TiO,MgI&5488&5421-5554 & FeI,MgI&8723&8677-8768\\  
  
CaI,FeI,TiO&5592&5554-5630 & FeI,MgI&8812&8768-8855\\  
\hline  
\end{tabular}  
\end{center}  
\vspace{-5mm}  
\end{table*}  
  
\section{Stellar population synthesis}  
The main purpose of a stellar population synthesis is to    
gather information on the stellar content of the galaxies   
under study; our synthesis also provides   
an evaluation of the dust content, the internal reddening   
being a parameter taken into account for the calculations.   
  
In addition to these pieces of information, the stellar   
synthesis also allow us to remove   
the starlight contribution from the data and consequently to   
unveil the optical emission-line features present in the   
spectra. Even though a few weak emission lines could be previously   
detected in some of the galaxies in our sample, their optical   
to near-UV spectra are clearly dominated by the stellar   
continuum. In fact, the fraction of starlight in   
such objects is so important that it almost completely swamps  
any strong emission feature associated with the active nucleus.  
  
The method used to compute the nuclear stellar populations and its   
application to our sub-sample of objects are described below.   
  
\subsection{The method}  
We have used a synthetic population algorithm developed by Pelat   
(1997). This algorithm uses a new mathematical method which gives a   
unique solution (Global Principal Geometrical solution, or GPG), contrary   
to the other methods widely used for population synthesis.  It uses the   
equivalent widths (EW) of all the absorption features found in the   
spectra. Essentially, it considers a galaxy as being    
made up of a set of stars with different spectral types, luminosities   
and metallicities. This particular composition will carry its own   
signature in terms of the EW of the absorption lines;   
it then defines the composition by matching this signature as closely   
as possible to the observed EW, using the least squares method.   
The internal reddening is also a free parameter given by the method   
in an indirect way.   
  
The accuracy of the fit is estimated by 
a parameter, the {\it distance}, which is defined as the sum, for   
all absorption features, of the difference between the observed EW   
and the synthetic EW calculated through the combination of the different   
stars that compose the stellar populations obtained. This value depends   
on the number of absorption features used and the smaller this   
{\it distance}, the better the fit. In addition, residuals estimated   
over the continuum help us to verify the accuracy of the solution found.   
The theoretical aspects of the method are described in Pelat (1997).   
This method was already applied to the synthesis of nuclear stellar   
populations in AGN and proved to give very good results (Serote Roos   
1996; Serote Roos et al. 1998; Boisson et al. 2000).   
  
The stars used for the synthesis were compiled from the Silva \& Cornell  
(1992) library. These authors have made available 72 stellar types with  
a large coverage in spectral types, luminosity classes and metallicities.  
The library wavelengths range from 3500~\AA\ to 9000~\AA, thus covering   
quite well the spectral range of our data. Their resolution   
being 11~\AA, we have convolved the stars in the library with a Gaussian   
with $\sigma$ = 7.1, in order to match the resolution of our spectra   
(20~\AA).   
The stellar library has been chosen in order to cover the space   
temperature/gravity as much as possible without being degenerate. In fact,   
we cannot include as many stellar types as we would like to, in order   
to prevent stellar library degeneracy, i.e. having two different   
stellar types with spectral energy distributions similar enough to be   
indistinguishable in a mathematical sense.  
Some super metal rich stars (SMR) have also been included. In total,   
the stellar library used contains 30 stars.  
  
We have identified and  
measured the EW of all absorption features ($\sim$\,80) present  
in the spectra, as well as the values of the continuum for each feature.  
Identification of the absorption lines and bands, together with the  
wavelength ranges used to measure the EW of the features, can be found  
in Table~\ref{EW}.  
  
\onecolumn  
\begin{sidewaystable}[!h]  
\centering  
\caption{\label{pop_synt_tab}  
Population synthesis solutions for each galaxy in our sample.}  
\begin{center}  
\begin{tabular}{lccccccccccccccccccc}  
\hline  
\hline  
Library & 0035 & 0055 & 0116 & 0149 & 0651 & 0729  
        & 0733 & 0848 & 0902 & 1144 & 1146 & 1241  
        & 1245 & 1558 & 1658 & 1703 & 1755 & 2202  
        & 2320 \\  
Stars & +227 & +300 & +319 & +710 & +410  & +562  
      & +597 & +686 & +468 & +352 & +596  & +735  
      & +676 & +595 & +302 & +223 & +626  & +363  
      & +203 \\  
\hline  
 O5V      &  0&  0&  0&  0&  0&  0&  0&  0&  0&  0  
          &  0&  0&  0&  0&  0&  0&  0&  0&  0\\  
 B3-4V    &  0&  0&  0&  0&  0&  0&  0&  0&  0&  0  
          &  0&  0&  0&  0&  0&  0&  0&  0&  0\\  
 A1-3V    &  0&  0&  0&  0&  0&  0&  0&  0&  0&  0  
          &  0&  0&  0&  0&  0&  0&  0&  0&  0\\  
 A8V      &  0&  0&  0&  0&  0&  0&  0&  0&  0&  0  
          &  0&  0&  0&  0&  0&  0&  0&  0&  0\\  
 F8-9V    &  0&  0&  0&  0&  0&  0&  0&  0&  0&  0  
          &  0&  0&  0&  0&  0&  0&  0&  0&  0\\  
 G9K0V    & 28& 63&  0& 29& 49&  0& 24&100& 87&  0  
          &  0&  0&  0&  0&  6&  0& 17&  0& 82\\  
 K5V      &  0&  0&  0&  0&  0&  0&  0&  0&  0&  0  
          &  0&  0&  0&  0&  0&  0&  0&  0&  0\\  
 M2V      &  2&  7&  0& 11&  0&  5&  2&  0&  0&  5  
          &  2&  9& 16&  4&  6&  7&  4&  3&  8\\  
 r F8V    &  0&  0&  0&  0&  0&  0&  0&  0&  0&  0  
          &  0&  0&  0&  0&  0&  0&  0&  0&  0\\  
 r G5V    &  0&  0& 33&  0&  0&  0&  0&  0&  0&  0  
          & 51&  0&  0&  0& 47&  0&  0&  0&  0\\  
 r K0V    & 20&  0&  0&  0&  0& 40& 35&  0&  1& 70  
          &  0&  0& 63& 43&  0& 24& 51&  0&  0\\  
          &   &   &   &   &   &   &   &   &   &  
          &   &   &   &   &   &   &   &   &   \\  
 O7B1III  &  0&  0&  0&  0&  0&  0&  0&  0&  0&  0  
          &  0&  0&  0&  0&  0&  0&  0&  0&  0\\  
 B5III    &  0&  0&  0&  0&  0&  0&  0&  0&  0&  0  
          &  0&  0&  0&  0&  0&  0&  0&  0&  0\\  
 B9III    &  0&  0&  0&  0&  0&  0&  0&  0&  0&  0  
          &  0&  0&  0&  0&  0&  0&  0&  0&  0\\  
 F4-7III  &  0&  0&  0& 32&  0&  0&  0&  0&  0& 17  
          &  0&  0&  0&  0&  0&  0&  0&  0&  0\\  
 G0-4III  &  0&  1& 19& 14&  0&  8&  0&  0&  0&  0  
          &  0& 39&  0&  0&  0&  0&  0& 47&  0\\  
 G8III   &  0&  0&  0&  0&  0&  0& 27&  0&  0&  0  
          & 36&  0&  0& 20&  0&  0&  9&  0&  0\\  
 K4III    &  0&  0&  0&  0&  0&  4&  0&  0&  0&  0  
          &  0&  0&  7&  4&  0&  0&  0&  0&  0\\  
 M0-1III   &  2&  0&  0&  0&  9&  5&  0&  0&  0&  0  
          &  4&  3&  0&  0&  0&  0&  0&  1&  0\\  
 M3III    &  0&  0&  0&  1&  0&  0&  0&  0&  0&  0  
          &  0&  0&  0&  0&  0&  0&  0&  0&  0\\  
 M5III    &  3&  0&  5&  0&  0&  0&  0&  0&  0&  0  
          &  0&  0&  0&  0&  0&  0&  0&  0&  0\\  
 r K3III  &  9& 11&  0&  0&  0&  0&  0&  0&  0&  0  
          &  0& 22&  0&  0&  0& 20&  8&  0&  0\\  
          &   &   &   &   &   &   &   &   &   &  
          &   &   &   &   &   &   &   &   &   \\  
 O8I      &  0&  0&  0&  0&  0&  0&  0&  0&  0&  0  
          &  0&  0&  0&  0&  0&  0&  0&  0&  0\\  
 B3-5I    &  0&  0&  0&  0&  0&  0&  0&  0&  0&  0  
          &  0&  0&  0&  0&  0&  0&  0&  0&  0\\  
 A7-9I    &  0&  0&  0&  0&  0&  3&  0&  0&  0&  0  
          &  0&  0&  0&  0&  0&  0&  0&  0&  0\\  
 G0-1I    & 35& 15& 24&  0&  0&  0&  0&  0&  0&  0  
          &  0& 21&  0&  0&  0& 45&  3& 25&  0\\  
 K1-2I    &  1&  0&  0& 13& 39& 15&  0&  0&  0&  7  
          &  0&  0&  0&  0& 26&  0&  0&  0&  6\\  
 K3-5I    &  0&  0&  0&  0&  0& 12& 10&  0&  7&  0  
          &  3&  0&  7& 29&  5&  3&  7& 19&  0\\  
 r K3I    &  0&  0& 12&  0&  0&  0&  0&  0&  0&  0  
          &  0&  2&  0&  0&  0&  0&  0&  0&  0\\  
 CO       &  0&  1&  6&  0&  2&  6&  2&  0&  4&  1  
          &  3&  3&  6&  0& 10&  0&  0&  5&  3\\  
\hline  
$distance$ & 53 & 43 & 115 & 157 & 83 & 50 & 153 & 89 & 64 & 70  
           & 111 & 70 & 253 & 147 & 64 & 88 & 93 & 87 & 102 \\  
lines used & 52 & 52 &  58 &  48 & 47 & 48 &  55 & 49 & 48 & 53  
           &  63 & 44 &  55 &  54 & 52 & 49 & 45 & 44 & 51 \\  
 $E(B-V)$  & 0.1 &0.15 &0.22 & 0.1 & 0.0 & 0.1 & 0.15 & 0.3 & 0.25 & 0.25  
           & 0.25 & 0.0 & 0.0 & 0.0 & 0.1 & 0.03 & 0.1 & 0.1 & 0.15 \\  
\hline  
\end{tabular}  
\end{center}  
\end{sidewaystable}  
\twocolumn  
\newpage  
  
The wavelength ranges have been defined taking into  
account the shape of the absorption features in both hot and cool stars  
of the stellar library. The continuum level has been determined globally over  
the whole wavelength range. The error due to the uncertainty on the continuum  
level is dominant over all other measurements and statistical uncertainties.  
For well defined strong features (e.g. Na\,ID) this error is always   
less or equal to 1~\AA\ in absolute value. It can go up to a few   
Angstr\"om for wide bands and strong blends.  
  
The EW of the same spectral features have been measured   
in the spectra of the observed galaxies.  
For these, we discarded the intervals corresponding to the emission  
lines, as well as those showing atmospheric absorption features.  
We would like to stress the   
fact that discarding some intervals does not affect the results of   
the method, as we still have  used a very large number of parameters   
(between 44 and 63 EW, depending on the objects).    
  
\subsection{Synthesis results}  
\subsubsection{Stellar content}  
Several studies performed in normal, nearby and distant ellipticals,  
in the field and in clusters  
(e.g. Pickles 1985; Carter et al. 1986;  
Couture \& Hardy 1988, 1990, 1993; Munn 1992; Davies et al. 1993;  
Brown 1996; Trager 1997; Charlot et al. 1998; Mobasher \& James, 2000;  
Minniti 2001) show old   
stellar populations, with a main contribution of late stellar types,   
typically main sequence stars. However, for the more luminous   
ellipticals a non-negligeable component of O-B main sequence stars   
seems to be present in many objects. The populations in the centre   
of the galaxies are generally super metal rich dwarf and/or subdwarf stars,   
contrary to the ones found in the bulge. A population gradient   
is usually found, with the populations being younger and richer   
in metals at the centre of the galaxies.   
Concerning galaxies in clusters, the younger populations of red giants  
and supergiant stars are usually found in the ellipticals located at   
the outskirts of the cluster.   
 
\begin{figure*}  
\begin{center}  
\resizebox{!}{23cm}{\includegraphics{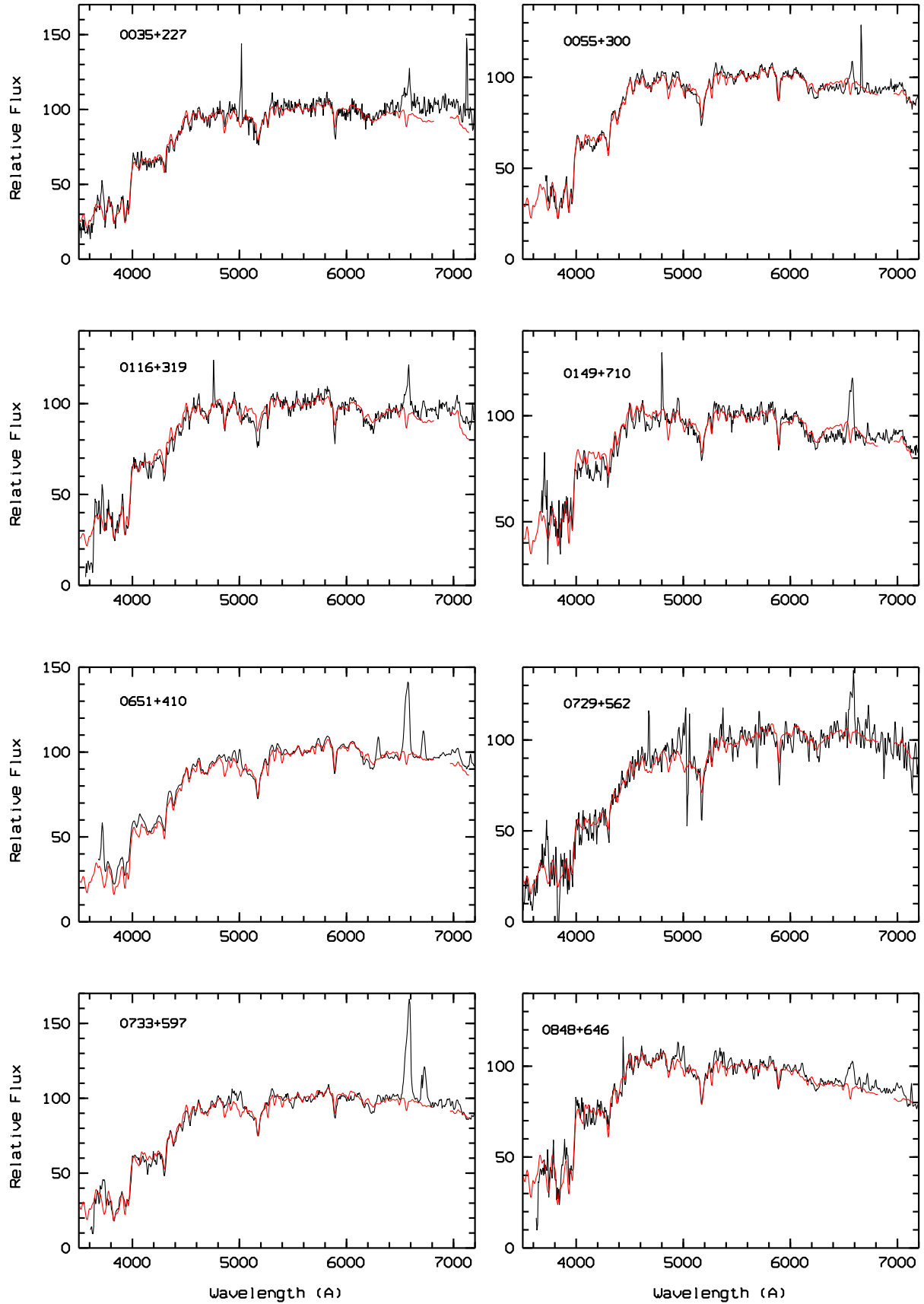}}  
\end{center}  
\caption{\label{obs_spectra}  
Observed and synthetic spectra in the rest frame. The observed   
spectra (in black) are superimposed on the synthetic ones   
(in gray).}   
\end{figure*}

In Fig.~\ref{obs_spectra} we plot the observed spectra (black lines)  
superposed on the synthetic ones (grey lines); all spectra are  
given in the rest frame.   
  
Table~\ref{pop_synt_tab} shows the stellar populations obtained   
for the 19 objects studied, together with the value of the   
{\it distance} for each solution, the number of absorption   
features used in the synthesis, and the internal reddening derived.  
  
Due to the nature of the synthesis method we use, it is important   
to have in mind several considerations when analysing the results   
obtained for each object. In particular, we do not pretend to derive   
the stellar content of the objects in its very details, but to   
have an idea of its main components in terms of hot and cold   
stars, young and old populations; note that the young population is   
given not only by hot stars, but also by supergiants.  
  
In a general way, we find old stellar populations typical of   
normal elliptical galaxies. For more than half of the objects   
in our sample, we do not find much dust. Concerning the   
super metal rich population, we have made some tests in order   
to estimate the importance of this component. Except for   
0116+319 and 1241+735, this population is not important; we   
find  good solutions to fit the spectra without it, as well.   
The hot stars are never present in the solutions, the   
populations being dominated by G, K and M spectral   
types. Thus, the younger populations appear in the form of   
supergiant stars (typical ages of the order of $10^7$ years).   
Despite supergiants being a very special class (which can   
include various types of stars, from the very massive and   
very young ones to the AGB or post-AGB stars, i.e. very   
old stars in their final phases), the fact that we are observing   
them in the optical range signifies that the contribution   
from supergiantes we are seeing   
comes from the young stars (Bruzual \& Charlot, 1993);   
in fact, AGB and post-AGB stars contribute  mainly in the UV range.    
  
For several objects in the sample (0733+597, 0848+686,   
1144+352, 1146+596 and 1755+626) we find only a small   
contribution of supergiants, implying that, in these objects,    
the population is quite old with ages of the order of   
$10^9$ -- $10^{10}$ years.   
For 0651+410, 0902+468, 1245+676, 1658+302 and 2320+203,   
the largest contribution comes from main sequence and supergiant   
stars, the contribution from giant stars being quite   
small ($\leq$ 10\%); this means that the old component comes   
from a non-evolved population of main sequence stars.   
For 0651+410 and 1658+302, a strong (41\%) young component    
in the form of supergiants appears.  
A lack in main sequence stars (contribution of $\leq$ 10\%)   
is found for two of the galaxies in the sample: 1241+735 and   
2202+363; in these cases,  the population is distributed more   
or less equally between giant and supergiant stars.   
  
NGC 3379, a normal elliptical was studied using the same method of  
computing the stellar populations (Boisson et al., 2000).  For this  
galaxy we have found a very homogeneous old population (with almost no  
population gradients in the central regions of the galaxy)   
with a main component of  
main sequence stars (K and M spectral types) and some contribution  
from giant (K, M) stars. This population seems to be supermetal rich,  
i.e. composed of stars with a metallicity higher than solar. This metal  
enrichment is a direct proof of strong star formation in the past.  
Little dust is present in the solutions.  
  
\addtocounter{figure}{-1}  
\begin{figure*}  
\begin{center}  
\resizebox{!}{23cm}{\includegraphics{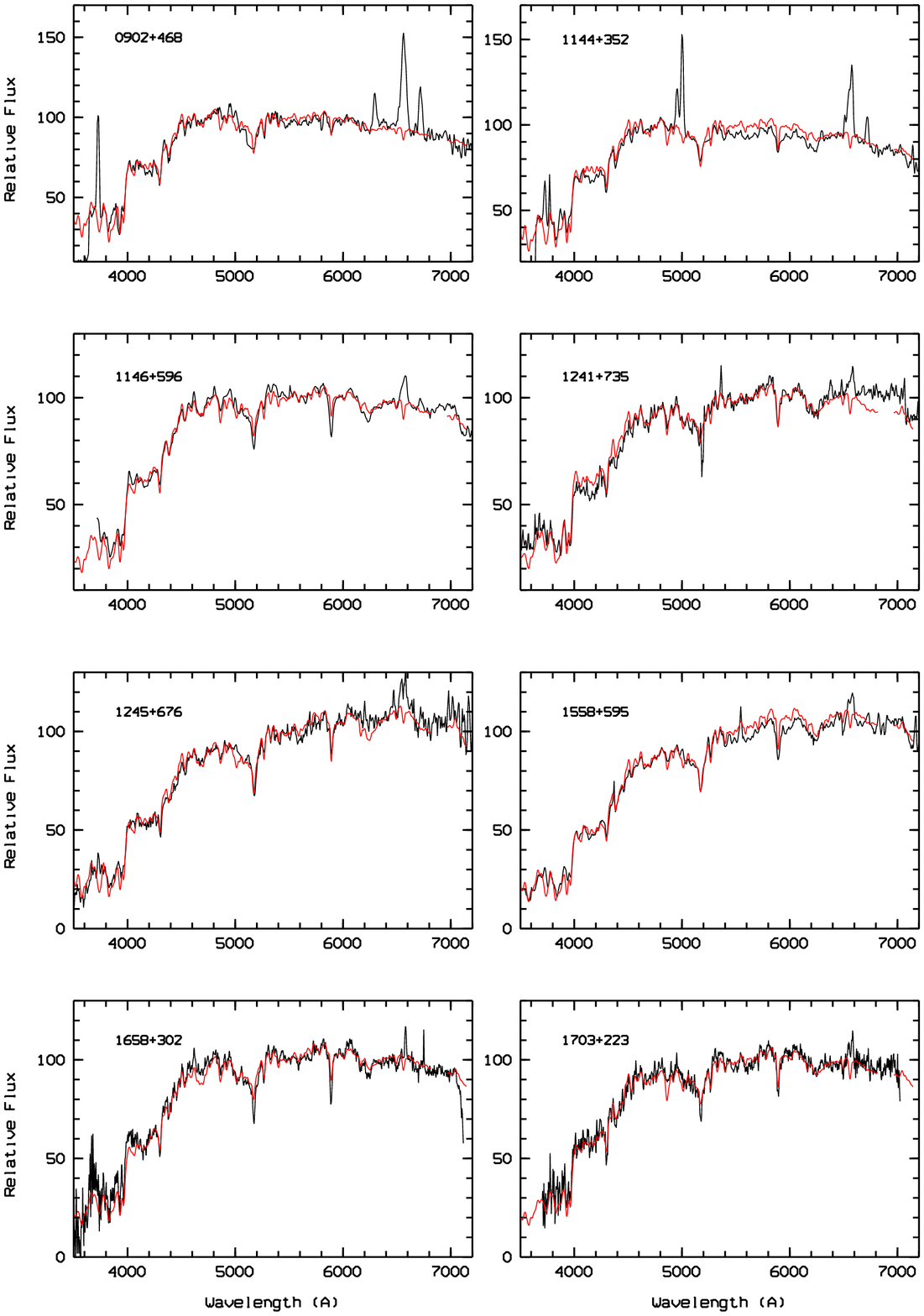}}  
\end{center}  
\caption{Observed and synthetic spectra (cont.).}  
\end{figure*}  
\addtocounter{figure}{-1}  
\begin{figure}  
\begin{center}  
\resizebox{8.5cm}{!}{\includegraphics{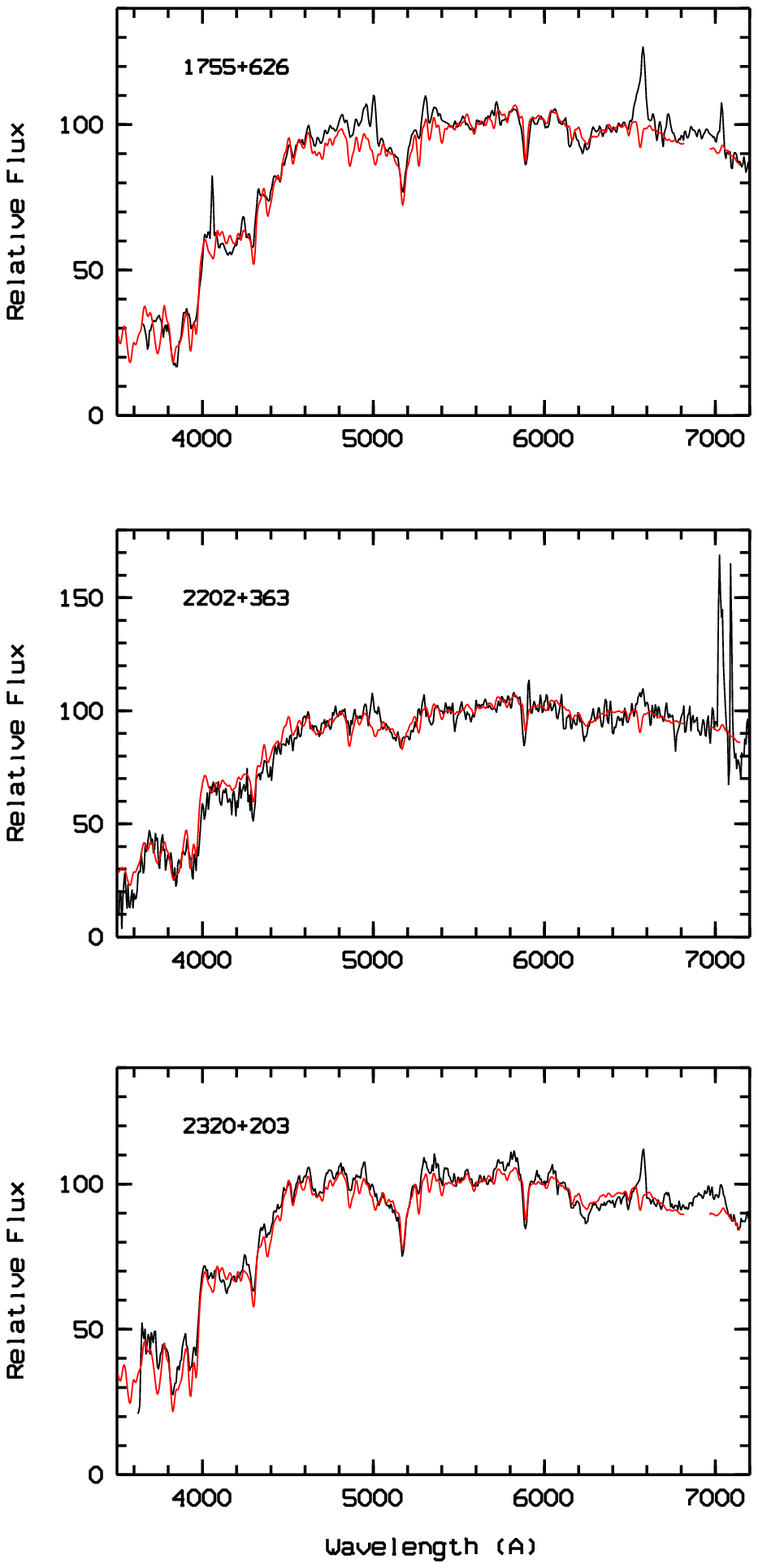}}  
\end{center}  
\caption{Observed and synthetic spectra (end). }  
\vspace{5mm}  
\end{figure}  
  
Comparing these results with the ones found in the present work, for  
our sample of galaxies, we verify that our objects have equally an old  
stellar population, the main contributors being G, K and M stars of  
luminosity classes V and III. The main difference is that   
our populations do not seem to be metal rich (except in the case   
of two objects: 0116+319 and 1241+735), contrary to the one found for   
NGC~3379.  This can be due to the fact that only the central regions of  
NGC~3379 are discussed in Boisson et al.'s study, whereas in this   
work we are dealing with an integreted spectrum of the whole galaxy.    
Since star formation can be really intense in the central regions,   
this metal enrichment would be more proeminent there, thus   
justifying the difference found.  
  
\subsubsection{Colours and ages}  
In order to have another constraint on the population ages,   
we have computed the $UBVRI$ colours of our objects. Only the   
$B-V$ and $V-R$ indicies were calculated; we did not use   
the $U$ and $I$ colours because these filters do not   
fall entirely in our wavelengh range, and therefore we would   
be losing too much light to achieve confident results.  
Using evolutionary synthesis models for elliptical  
galaxies, from the Bruzual \& Charlot code (Bruzual et al. 1997), we have   
estimated the ages of the populations.  
Two different kinds of models were used: a single stellar population model,  
i.e. a model computing the spectral evolution of a single initial   
instantaneous-burst of star formation,   
and a model that computes the evolution of the stellar populations  
with longer time scales of star formation, i.e.  
where the star formation rate (SFR) decreases exponentially with  
time (that is with the available gas content), during 1 Gyr.  
For both models the results are essentially the same.   
We find that the colour indices of most spectra are compatible with an   
intermediate/old stellar population with ages ranging from   
$10^9$ to $10^{10}$ years. This is in  agreement with the   
results of our population synthesis, corroborating the method   
and our conclusions.   
  
\subsubsection{Dust content}  
A degenerency exists between the galaxy's internal   
dust and the blue continuum of the spectrum, i.e. we can fit   
the observed continnum either with a given amount of dust   
plus a given blue slope, or just with a certain amount   
of dust  (smaller in this case). This happens because both   
contributions have the effect of blueing the spectrum. In this   
work we have fitted all spectra in such a way that the continuum   
is also perfectly fitted by a value of $E(B-V)$, thus assuming   
that all contribution to the continuum comes from the dust and   
stellar content of the host galaxy. Instead, we could have   
found a blue excess in the continuum by adding more dust to   
the solutions and then argue about this quantity coming from the   
active nucleus.  However, in this case it would be difficult to  
quantify either the quantity of dust present in the object, or the  
slope of the blue continuum.  Also, and from the values of the Balmer break  
contrasts of the objects, we estimate the non-stellar continuum to be  
negligeable, well within the error bars of the flux calibration of the  
spectra. This is why we chose not to include   
a contribution from an AGN to the continuum.   
  
\section{Summary and Conclusions}  
  
In an effort to better understand LL FRS sources and their   
host galaxies, we undertook the spectroscopic study of 19   
objects selected from March\~a's et al. 200 mJy sample;   
this was known to contain a high fration of objects oriented   
close to the line-of-sight, such as LL BLLs. This study focused   
on the stellar content of the host galaxies.  
  
We performed stellar population synthesis aimed at identifying   
the nuclear stellar populations present in the objects; our work   
made use of a very reliable mathematical method, yielding a   
Global Principal Geometrical solution.   
  
Our main result  concerns the stellar populations found in these    
objects, which are in agreement with those found in elliptical   
galaxies; our results show that the nuclear populations are   
composed of old stars of solar metallicity or   
lower; the dust content is weak, which is also typical of ``normal''   
elliptical galaxies.  
It is most interesting to note that similar stellar populations   
were reported in Low Ionization Nuclear Emission-Line Regions. In   
particular,  Serote Roos (1996) and Boisson et al. (2000) have shown,   
in a study of the stellar populations in a sample of AGN with different   
types of activity, that LINERs present the more evolved  populations of   
the sample, with little recent star formation,  i.e. the central   
regions of these objects are largely dominated by a red evolved    
population of metal rich stars.   
  
The present work demonstrates the importance of   
stellar population synthesis in the study of low luminosity   
radio sources. We have shown that stellar synthesis allows us to   
obtain information on the stellar populations   
of the host galaxies, therefore providing material for   
further studies on the connection between host galaxy and the active   
nucleus. Stellar synthesis also constitutes a precious tool   
allowing us to unveil the optical emission lines present   
in the spectra of our objects. The analysis of such features    
provides information on the nature of the nuclear emission regions   
and excitation mechanisms at work in LL FSR sources; such a study   
has been performed and is the subject of a companion paper (Paper~II).   
  
\bigskip  
{\it Acknowledgements:}  
We are greatly indebted to D. Pelat for the use of the  
population synthesis GPG program. We thank  
Thibault Lejeune for kindly computing the colours for our objects as well  
as the evolutionary synthesis models used in this work.  
We thank M. March\~a and   
A. Caccianica for fruitful discussions, and P. V\'eron for valuable   
comments. M. Serote Roos acknowledges financial  
support from FCT, Portugal, under grants no. BPD/9995/96 and  
BPD/5684/01, and from  
FCT/ESO Project ref. PESO/C/PRO/1254/98.  A.\,C. Gon\c{c}alves   
acknowledges financial support from FCT, Portugal, under grant no.    
SFRH/BPD/9422/2002.    
  
  
\end{document}